\documentclass[12pt,a4paper]{article}
\usepackage{amssymb,epsfig,amsmath,latexsym}
\newtheorem{theorem}{Theorem}[section]
\newtheorem{proposition}[theorem]{Proposition}
\newtheorem{definition}[theorem]{Definition}

\newtheorem{corollary}[theorem]{Corollary}

\newcommand{\pf}{\noindent{\textit{Proof:} }}
\newcommand\Q[1]{$#1$ \quad}

\newcommand{\TT}{\mathbb{T}}
\newcommand{\ZZ}{\mathbb{Z}}
\newcommand{\FF}{\mathbb{F}}
\newcommand{\RR}{\mathbb{R}}
\newcommand{\CC}{\mathbb{C}}

\begin{document}
\title{Quaternions over Galois rings and their codes}

\author{Pierre Lance A. Tan\thanks{Corresponding Author} \: \: and \: Virgilio P. Sison\\ Institute of Mathematical Sciences and Physics \\ University of the Philippines Los Ba\~{n}os \\ College 4031, Laguna, Philippines \\ Emails: \{patan3, vpsison\}@up.edu.ph \\ \\} 

\maketitle

\begin{abstract}
It is shown in this paper that, if $R$ is a Frobenius ring, then the quaternion ring $\mathcal{H}_{a,b}(R)$ is a Frobenius ring for all units $a,b \in R$. In particular, if $q$ is an odd prime power then $\mathcal{H}_{a,b}(\FF_q)$ is the semisimple non-commutative matrix ring $M_2(\FF_q)$. Consequently, a homogeneous weight that depends on the field size $q$ is obtained. On the other hand, the homogeneous weight of a finite Frobenius ring with a unique minimal ideal is derived in terms of the size of the ideal. This is illustrated by the quaternions over the Galois ring $GR(2^r,m)$. Finally, one-sided linear block codes over the quaternions over Galois rings are constructed, and certain bounds on the homogeneous distance of the images of these codes are proved. These bounds are based on the Hamming distance of the quaternion code and the parameters of the Galois ring. Good examples of one-sided rate-$2/6$, $3$-quasi-cyclic quaternion codes and their images are generated. One of these codes meets the Singleton bound and is therefore a maximum distance separable code. 

\end{abstract}

\noindent {\it Keywords}: quaternions, Galois rings, Frobenius rings, homogeneous weight\\
\noindent {\it MSC 2020}: 16H05, 16L99, 94B05, 94B65

\section{Introduction}
The classical notion of quaternions began with Hamilton in 1843 when he attempted to describe three-dimensional problems in mechanics. Much like how the field of complex numbers $\CC$ is an extension of the field $\RR$ of real numbers, so too is the ring of real quaternions which forms a $4$-dimensional central simple algebra over $\RR$ with basis $\{1,i,j,k\}$ such that $i^2=j^2=k^2=ijk=-1$, but with the commutative law of multiplication removed. The quaternions over $\RR$ are the only noncommutative division ring over the real numbers. They can be viewed as $2 \times 2$ complex matrices of the form \begin{equation}\label{matrix}\left(\begin{matrix} z & w \\ -\overline{w} & \overline{z} \end{matrix}\right).\end{equation}      

Quaternions have seen applications in computer science and other fields of mathematics. The multiplication rules of quaternions make algorithms faster and more compact than when matrix representations are used.

More recently, quaternion algebras have been studied as Frobenius algebras \cite{lee}. An algebra $A$ over a commutative ring $K$ with unity is a ring that behaves as a unitary (left) module over $K$ such that $k(ab)=(ka)b=a(kb)$ for all $k \in K$ and $a,b \in A$. Moreover, quaternions have been constructed over finite fields including the prime field $\FF_p$ \cite{miguel}. The quaternions over $\FF_p$, where $p$ is odd, are simply the matrix ring $M_2(\FF_p)$, which has been the subject of study in \cite{corro} in the context of orbit codes. This is further extended to quaternions over finite commutative rings with unity, such as the residue class ring $\ZZ_n$ of integers modulo $n$ \cite{grau}.

Quaternions have been further generalized in numerous directions. There is the octonion algebra, which is an 8-dimensional extension of $\RR$.  Using the Cayley-Dickson construction as given in \cite{dickson}, the dimension of the quaternions is doubled, however the property of associativity is lost.

It is quite natural to deal with quaternions over finite Frobenius rings which include not only $\ZZ_n$, but more generally all the Galois rings, such as the finite fields and the ring $\ZZ_{p^r}$ where $p$ is prime and $r$ is a positive integer. This paper shows that quaternions over Frobenius rings, in particular, over Galois rings, are Frobenius rings on which a suitable homogeneous weight can be defined. Then the so-called quaternion codes that satisfy certain good distance properties are constructed.

The material is organized as follows: Section~\ref{sect:prelim} provides the conceptual framework while Section~\ref{sect:results} presents the main results.

\section{Preliminaries and definitions}
\label{sect:prelim}

This section tackles briefly the structural properties of Galois rings, Frobenius rings, quaternion rings and linear block codes. For a more detailed discussion of these topics, we refer the reader to \cite{grau}, \cite{honold} and \cite{wood}.

\subsection{Galois rings}
Let $p$ be a prime number and $r\geq 1$ an integer. Consider the finite commutative ring $\ZZ_{p^r}$ of integers modulo $p^r$. When $r=1$ then the ring $\ZZ_p$ is a field and is usually denoted by $\FF_p$. Let $\ZZ_{p^r}[x]$ be the ring of polynomials in the indeterminate $x$ with coefficients in $\ZZ_{p^r}$. The Galois ring with characteristic $p^r$ and cardinality $p^{rm}$ is the residue class ring $\ZZ_{p^r}[x]/(h(x))$, where $h(x)$ is a monic polynomial of degree $m$ in $\ZZ_{p^r}$ whose image under the mod $p$ reduction map is an irreducible polynomial in $\FF_p[x]$. The Galois ring can also be seen as a Galois extension $\ZZ_{p^r}[\omega]$ of $\ZZ_{p^r}$ by a root $\omega$ of $h(x)$. Thus, every element has a unique additive representation
\begin{equation}
x=b_0+b_1\omega+b_2\omega^2+\dots+b_{m-1}\omega^{m-1}
\end{equation}
where $b_i \in \ZZ_{p^r}, i=0,1,\dots,m-1$, which makes it a free module over $\ZZ_{p^r}$ with basis $\{1,\omega,\omega^2,\dots,\omega^{m-1}\}$. The ring satisfies the invariant dimension property in the sense that any other basis has exactly $m$ elements.

Since any two Galois rings of the same characteristic and the same cardinality are isomorphic, we shall use $GR(p^r,m)$ to denote such rings. When $m=1$, the Galois ring $GR(p^r,m)$ is the integer ring $\ZZ_{p^r}$, and when $r=1$, we obtain the Galois field $\FF_{p^m}$. The ring $GR(p^r,m)$ is a commutative local principal ideal ring and is a finite chain ring of length $r$ with its unique maximal ideal $(p)$ containing the nilpotent elements.

\subsection{Frobenius rings}
Let $R$ be a finite ring with unity, and $\TT$ be the multiplicative group of unit complex numbers. The group $\TT$ is a one-dimensional torus. A character of $R$, considered as an additive abelian group, is a group homomorphism $\chi:R\rightarrow \TT$. A character $\chi$ of $R$ is called a left (resp. right) generating character if and only if $\ker \chi$ contains no nonzero left (resp. right) ideals. The ring $R$ is called Frobenius if and only if $R$ admits a left or a right generating character. A character $\chi$ on $R$ is a left generating character if and only if it is a right generating character \cite{wood}. Alternatively, a finite ring $R$ is Frobenius if and only if the left (resp. right) socle of $R$ is a left (resp. right) principal ideal.

We note that any finite principal ideal ring is Frobenius. In particular, the Galois ring $GR(p^r,m)$ is a Frobenius ring with generating character
\begin{equation}\label{galois}
\chi(x)=e^{2\pi ib_{m-1}/p^r} \text{ for } x=\Sigma ^{m-1}_{i=0} b_i \omega^i
\end{equation}
Furthermore the non-commutative ring $M_n(R)$ of $n \times n$ matrices over a Frobenius ring $R$ is Frobenius with generating character \begin{equation} \chi^*(B) = \chi(tr(B)) \end{equation} where $\chi$ is a generating character of $R$ and $tr$ is the usual matrix trace.
\subsection{Homogeneous weight}
A left homogeneous weight on an arbitrary finite ring $R$ with unity is defined in the sense of \cite{greferath2000} in terms of the nonzero principal left ideals. Right homogeneous weights are defined analogously.

\begin{definition}
A weight function $w:R\rightarrow\RR$ on a finite ring $R$ is called (left) homogeneous if $w(0)=0$ and the following are true.
\begin{enumerate}
	\item If $Rx=Ry$, then $w(x)=w(y)$ for all $x,y\in R$.
	\item There exists a real number $\Gamma\geq 0$ such that $\dfrac{\Sigma_{y\in Rx}w(y)}{|Rx|}=\Gamma$, for all nonzero ideals $Rx$.
\end{enumerate}
\end{definition}

If a weight is both left homogeneous and right homogeneous, we call it simply as a homogeneous weight. The constant $\Gamma$ is called the average value. The homogeneous weight $w$ can be normalized by replacing it with $\widetilde{w}=\Gamma^{-1}w$. It was proved in \cite{honold} that, if $R$ is Frobenius with generating character $\chi$, then every homogeneous weight $w$ on $R$ can be expressed in terms of $\chi$ as
\begin{equation}
w(x)= \Gamma \left[ 1-\dfrac{1}{|R^{\times}|}\sum_{u\in R^{\times}}\chi(xu) \right]
\end{equation}

\noindent where $R^{\times}$ is the multiplicative group $R$.

Another formula for the homogeneous weight is given in terms of the M\"{o}bius function $\mu$ on the poset of the principal left ideals of $R$ with the property that (i) $\mu(Rx,Rx)=1$ for all $x\in R$, (ii) $\mu(Ry,Rx)=0$ if $Ry \nleq Rx$, and (iii) $\Sigma_{Ry\leq Rz \leq Rx} \mu(Rz,Rx)=0$ if $Ry\leq Rx$.

\begin{equation}\label{mobius}
w(x)=\Gamma \left[ 1 - \dfrac{\mu(0,Rx)}{|R^{\times} x|} \right]
\end{equation}

The homogeneous weight of $GR(p^r,m)$ as a finite chain ring is derived in \cite{greferath2001} as follows.
\begin{equation}\label{hom}
w_{\textnormal{hom}}(x)=
\begin{cases}
0, & \text{if } x=0 \\
p^{m(r-1)}, & \text{if } x\in (p^{r-1}) \setminus \{0\}\\
(p^m-1)p^{m(r-2)}, & \text{otherwise}
\end{cases}
\end{equation}
in which the average value $\Gamma$ is computed as follows.
\begin{equation}\label{gamma}
\Gamma = (p^m-1)p^{m(r-2)}
\end{equation}
This value happens to be the minimum nonzero weight, while the maximum weight is $p^{m(r-1)}$.

\subsection{Quaternion rings}

Let $R$ be a commutative ring with unity and $a,b$ are units. The quaternion ring over $R$ denoted by $\mathcal{H}_{a,b}(R)$ consists of elements of the form
\begin{equation}
x_0+x_1i+x_2j+x_3k 
\end{equation}
where $x_i \in R, i^2=a, j^2=b, ij=-ji=k$. Clearly, $\mathcal{H}_{a,b}(R)$ is a non-commutative ring with unity and a two-sided module over $R$ with basis $\{1,i,j,k\}$. Specifically, when $a=b=-1$, we denote the ring by $\mathcal{H}(R)$ in reference to Hamilton's quaternions. On the other hand, when $a=b=1$, we denote the ring by $\mathcal{L}(R)$ which is the convention used in \cite{grau}.

In the classical sense, quaternions have been defined over finite or infinite fields. At present there has been keen interest on quaternion rings over finite rings such as the residue class ring $\ZZ_n$, as shown in the following theorems.

\begin{theorem}
\textnormal{\cite{grau}} Let $R$ be the residue class ring $\ZZ_n$ and $M_2(R)$ the non-commutative ring of $2 \times 2$ matrices over $R$. Then $\mathcal{H}_{a,b}(R)\cong \mathcal{H}(R)$ if and only if $a\equiv b \equiv -1 \mod 4$, otherwise $\mathcal{H}_{a,b}(R)\cong \mathcal{L}(R)$. Moreover, $\mathcal{H}_{a,b}(R)\cong M_2(R)$ if and only if $n$ is odd.
\end{theorem}

\begin{corollary}
\textnormal{\cite{grau}} $\mathcal{H}_{a,b}(\ZZ_n)\cong \mathcal{H}(\ZZ_n) \cong \mathcal{L}(\ZZ_n)$ if and only if $n$ is odd or $n=2$.
\end{corollary}

\subsection{One-sided block codes}

A block code $C$ of length $n$ over an arbitrary finite ring $R$ with unity is a nonempty subset of the free module $R^n$ which contains all the $n$-tuples over $R$. The code $C$ is said to be left (resp. right) $R$-linear if $C$ is a left (resp. right) $R$-submodule of $R^n$. If $C$ is both left $R$-linear and right $R$-linear, we simply call $C$ a linear block code over $R$. If $C$ is a free module, then $C$ is said to be a free code. A $k\times n$ matrix $G$ over $R$ is called a generator matrix if its rows span $C$ and no proper subset of the rows generates $C$, that is $C=\{v\in R^n : v=uG, u \in R^k\}$ so that the rate of $C$ is $k/n$. In this paper, a block code over a quaternion ring will be called a {\it quaternion code}.

Let $w$ be a weight function on the ring $R$ which is extended coordinate-wise to $R^n$. The minimum weight of $C$ is the minimum of the weights of the nonzero elements of $C$. The distance metric $d:R^n \times R^n \rightarrow \RR$ is defined by $d(x,y)=w(x-y)$ for $x,y\in R^n$. The minimum distance $d$ of the code $C$  with respect to $w$ is defined to be $d=\min \{ d(x,y)|x,y \in C, x \neq y \}$. For linear codes, the minimum distance is equal to the minimum weight.

It is given in \cite{macwilliams} that the Singleton bound for codes over any alphabet of size $m$ is as follows.
\begin{equation}\label{singleton}
d_H \leq n - \log_m(|C|)+1
\end{equation}
where $d_H$ is the minimum Hamming distance of $C$.

Let $C$ be a linear block code of length $n$. The code $C$ is called cyclic if, for every codeword $c=(c_1,\dots,c_n)$ from $C$, the word $(c_n,c_1,\dots,c_{n-1})$ is again a codeword. The code $C$ is called $\ell$-quasi-cyclic if, a cyclic shift of a codeword by $q$ positions
results in another codeword, that is, for every codeword $c=(c_1,\dots,c_n)$ from $C$, the word $(c_{n-\ell+1},\dots,c_n,c_1,\dots,c_{n-\ell})$ is again a codeword. A $1$-quasi cyclic code is a cyclic code. 

\section{Results and discussion}\label{sect:results}

The results are divided as follows: First, we develop the theory of quaternions over Galois rings. There are several existing literature on quaternions over specific classes of Galois rings, but now we are able to unify these results. We claim that they are Frobenius rings and hence a homogeneous weight can be defined. Second, we construct the so-called quaternion codes and get their images as codes over Galois rings with four times the length. Theoretical distance bounds involving these codes are derived. Finally, new good examples of such codes and their distances are presented, one of which satisfies the Singleton bound.

\subsection{Quaternions over Galois rings}\label{sect:results1}

It has been proved in \cite{lee} that quaternions over Frobenius algebras are also Frobenius algebras. We shall prove it for the case of rings instead of algebras, which was a conjecture in the said paper. It follows that the rings are Artinian, finite and with unity. It is also clear that quaternions are modular with the standard basis $\{1,i,j,k\}$.

\begin{theorem}
Let $R$ be a Frobenius ring and $a,b \in R^{\times}$. Then the quaternion ring $\mathcal{H}_{a,b}(R)$ is also a Frobenius ring.
\end{theorem}

\pf Let $x=x_0+x_1i+x_2j+x_3k$ and $y=y_0+y_1i+y_2j+y_3k$. Let $\chi$ be a generating character of $R$. Define the map $\chi^*(x)=\chi(x_0)$. Then $\chi^*(x+y)=\chi(x_0+y_0)=\chi(x_0)\chi(y_0)=\chi^*(x)\chi^*(y)$. Thus $\chi^*$ is a group homomorphism and hence is a character of $\mathcal{H}_{a,b}(R)$. Furthermore, $\ker \chi^*=\{x:x_0\in \ker \chi\}$. We show that $\ker \chi^*$ contains no nonzero left ideals. Let $I\subseteq \ker \chi^*$ and $x\in I$. First observe that $x_0=0$, for if $Rx_0\neq\{0\}$ then $\ker \chi$ contains a nonzero left ideal $Rx_0$, which is a contradiction. Next observe that $x=0$, for if say $x_1\neq0$ then since $a$ is a unit, the scalar term of $ix$ is $x_1a\neq0$, which is a contradiction. Hence $\chi^*$ is a generating character. \Q{\Box}

If $R$ is finite then $\mathcal{H}_{a,b}(R)$ is finite with cardinality $|R|^4$. The following more specific result, which will be useful in the succeeding discussion, is an immediate consequence of the preceding theorem via (\ref{galois}).

\begin{corollary}
Let $R=GR(p^r,m)$ be a Galois ring. Then the quaternion ring $\mathcal{H}_{a,b}(R)$ is Frobenius.
\end{corollary}

It should be noted that, in a more general sense as given in \cite{grau}, if $R$ is a commutative, associative and unital ring, just like $GR(p^r,m)$, the quaternion ring $\mathcal{H}_{-1,-1}(R)$ is isomorphic to a subring of complex matrices given similarly in (\ref{matrix}). However, in a more specific case, the paper by Miguel \cite{miguel} proved that quaternion rings over the prime field $\FF_p$, where $p$ is odd, is isomorphic to the full matrix ring $M_2(\FF_p)$. A more general theorem is stated as follows.

\begin{theorem}
\textnormal{\cite{miguel}} Over a field $F$ with characteristic different from $2$, there are exactly two quaternion rings: the matrix ring $M_2(F)$ and a division ring.
\end{theorem}

This theorem leads us to extend the isomorphism as follows.

\begin{corollary}\label{m2}
If $p$ is an odd prime, then $\mathcal{H}_{a,b}(\FF_{p^m})\cong M_2(\FF_{p^m})$.
\end{corollary}

Since the choice of $a$ and $b$ no longer matters when $p$ is odd, we can simply use the notation $\mathcal{H}(\FF_q)$, where $q=p^m$. Because of Corollary~\ref{m2}, our study of quaternions now has access to the deep results in \cite{corro} on rings of matrices. We now combine the results of \cite{miguel}, which is limited to $\FF_p$, and \cite{corro}, which makes no reference to quaternions. Hence, the formulas for the number of zero divisors and idempotents are easily computed. These formulas coincide with the results in \cite{corro} for the matrix ring $M_2(\FF_q)$, where $q$ is a power of a prime.

\begin{corollary}
Let $\mathcal{H}(\FF_q)$ be the ring of quaternions over the finite field $\FF_q$ where $q=p^r$ (odd $p$). Then $\mathcal{H}(\FF_q)$ has $q^3+q^2-q-1$ zero divisors and $q^2+q+2$ idempotents.
\end{corollary}

The zero divisors and idempotents of $\mathcal{H}(\FF_q)$ each play a role in the homogeneous weight of $\mathcal{H}(\FF_q)$. The idempotents are the generators of the nontrivial proper ideals of $\mathcal{H}(\FF_q)$, which altogether contain all of the zero divisors. This is why the nonzero values of the homogeneous weight function, which we will see later, are segregated into the zero divisors and units.

The next results follow immediately from the work in \cite{corro} which uses (\ref{mobius}).

\begin{corollary}\label{ideals}
Let $\mathcal{H}(\FF_q)$ be the ring of quaternions over the finite field $\FF_q$ where $q=p^r$ (odd $p$). Then $\mathcal{H}(\FF_q)$ is a semisimple ring with $q+1$ proper left (resp. right) ideals, and moreover, each proper left (resp. right) ideal is principal,  minimal and maximal, and has $q^2$ elements.
\end{corollary}

\begin{corollary}\label{corroweight}
Let $\mathcal{H}(\FF_q)$ be the ring of quaternions over the finite field $\FF_q$ where $q=p^r$ (odd $p$). Then the homogeneous weight of $\mathcal{H}(\FF_q)$ with average value $\Gamma$ is given by
\[ w(x)=
\begin{cases}
0, & \text{if } x=0 \\
\Gamma\left(\dfrac{q^2}{q^2-1}\right), & \text{if } x \text{ is a zero divisor} \\
\Gamma\left(1-\dfrac{1}{(q-1)(q^2-1)}\right), & \text{if } x \text{ is a unit}
\end{cases}
\]
\end{corollary}

We highlight the fact that although these results speak of quaternions, it is only the field size $q$ that determines the above weights.

Since much work has been done on the case when $p$ is odd, we are interested in completing the case when $p$ is even. Note that the underlying Galois ring may have characteristic equal to $2$, which leads to a commutative quaternion ring. We discovered that the corresponding lattice contains a unique minimal ideal, from which the homogeneous weight is completely determined. By minimal left (resp. right) ideal $I$ we mean, $I \ne \{0\}$, and if $K$ is a left (resp. right) ideal with $\{0\} \subseteq K \subseteq I$, then either $K=\{0\}$ or $K=I$. 

\begin{theorem}\label{minimal}
Let $R$ be a finite Frobenius ring with a unique minimal left (resp. right) ideal $I$. The homogeneous weight on $R$ with average value $\Gamma$ is given as follows.
\[ w_{\textnormal{hom}}(x)=
\begin{cases}
0, & \text{if } x=0 \\
\Gamma\dfrac{|I|}{|I|-1}, & \text{if } x\in I \setminus \{0\}\\
\Gamma, & \text{otherwise}
\end{cases}
\]
\end{theorem}

\pf In the following arguments, we assume that all ideals are left ideals. The proof is analogous for right ideals. Let $x\in I$ be nonzero. Then $Rx\subseteq I$ and $Rx$ is nontrivial. But since $I$ is a minimal ideal, then $Rx=I$. This is true for all $x\in I$ nonzero. Therefore, all the $|I|-1$ nonzero elements have the same weight $W$. Furthermore, the average value is $\Gamma=W\dfrac{|I|-1}{|I|}$. Thus, we have proven the first statement. We remark that this part of the proof is applicable to rings that have more than one minimal ideal. Now, let $J$ be an ideal containing $I$ such that there is no proper ideal between $I$ and $J$. Then the elements of $J-I$ are associates. Since the average value across $I$ is already $\Gamma$, then by arithmetic computation, the weight of any element in $J-I$ must be $W=\Gamma\dfrac{|J-I|}{|J-I|}=\Gamma$. We can repeat this argument for all ideals $J_n$ satisfying the assumption. Thus, we know the weight of ideals one level above $I$. Furthermore, we can generalize the argument for each level of ideals by induction. We stress that this is due to the uniqueness of $I$. We can repeat this argument up to the whole ring, since it is finite. Hence, the weight of any element not in $I$ is $\Gamma$. \Q{\Box}


Note that the average value $\Gamma$ is still the minimum nonzero weight. A special case of Theorem \ref{minimal} occurs when $R$ is a finite chain ring. In this case, we know the exact value of $\Gamma$. For example, if $R=\FF_{p^m}$ then $I=R$ and

\begin{equation}
\Gamma\dfrac{|I|}{|I|-1} = \left(\dfrac{p^m-1}{p^m}\right) \left(\dfrac{p^m}{p^m-1}\right)=1.
\end{equation}
Therefore, the homogeneous weight is just the Hamming weight. On the other hand, if $R=\ZZ_{p^r}$ then $I=(p^{r-1})$ and

\begin{equation}
\Gamma\dfrac{|I|}{|I|-1} = (p-1)p^{r-2} \left( \dfrac{p}{p-1} \right) = p^{r-1}.
\end{equation}
Hence,
\begin{equation}
w_{\textnormal{hom}}(x)=
\begin{cases}
0, & \text{if } x=0 \\
p^{r-1}, & \text{if } x\in (p^{r-1}) \setminus \{0\}\\
(p-1)p^{r-2}, & \text{otherwise}
\end{cases}
\end{equation}
In general, it can be easily shown that applying Theorem \ref{minimal} to the Galois rings $GR(p^r,m)$ will give us the homogeneous weight in (\ref{hom}) with the unique minimal ideal $(p^{r-1})$ having $p^m$ elements. 

However, Theorem \ref{minimal} applies to more than just Galois rings and finite chain rings. In the succeeding discussion, we will cover $\mathcal{H}(\FF_{2^m})$ and $\mathcal{H}(\ZZ_{2^r})$, in which we will find explicit forms for the minimal ideal. This in stark contrast to $\mathcal{H}(\FF_{p^r})$ (odd $p$), where there is more than one minimal ideal. We begin by splitting the cases $r=1,m>1$ and $r>1,m=1$. We then generalize to $r>1,m>1$ from the latter.

\begin{theorem}\label{field}
The two-sided principal ideal $(x)=x\FF_{2^m}$, where
\begin{equation}
x=1+i+j+k
\end{equation}
is the only minimal ideal of $\mathcal{H}(\FF_{2^m})$. Furthermore, $|(x)|=2^m$.
\end{theorem}
\pf Let $y=y_0+y_1i+y_2j+y_3k$. Since $char(\FF_{2^m})=2$, positives and negatives are the same. Then $yx=xy=xz$, where $z=y_0+y_1+y_2+y_3\in\FF_{2^m}$. Thus $(x)\subseteq x\FF_{2^m}$. On the other hand, $\FF_{2^m}\subseteq \mathcal{H}(\FF_{2^m})$, so that $x\FF_{2^m}\subseteq(x)$. Hence, $(x)=x\FF_{2^m}$. Furthermore, it is easy to see that for any nonzero ideal $I$ with generator $y$, there is always a quaternion such that when multiplied to $y$, the product is a nonzero scalar multiple of $x$. Therefore $(x)$ is always contained in $I$. \Q{\Box}

With this, we have settled the case $r=1$. We separate this result from the cases $r>1$ because in those cases, the minimal ideal degenerates to two elements.

\begin{theorem}\label{ring}
The two-sided principal ideal $(x)=\{0,x\}$, where
\begin{equation}
x=2^{r-1}(1+i+j+k)
\end{equation}
is the only minimal ideal of $\mathcal{H}(\ZZ_{2^r})$.
\end{theorem}
\pf The proof is analogous to that of Theorem~\ref{field}, as evident in the $1+i+j+k$ term and due to $2^{r-1}$ having characteristic $2$. The distinction lies in the fact that $\FF_{2^m}$ is a field, hence the minimal ideal of $\mathcal{H}(\FF_{2^m})$ has cardinality $2^m$. In contrast, $2^r$ is a zero divisor that, when multiplied to any other scalar, results in $2^r$ or $0$. \Q{\Box}

We remark that from Corollary~\ref{ideals}, Theorem~\ref{field} and Theorem~\ref{ring}, we can give another proof that such rings are Frobenius. We reason with the socle definition of Frobenius rings. In the case of Corollary~\ref{ideals}, the socle is the entire ring, while in the case of Theorem~\ref{field} (resp. Theorem~\ref{ring}), the socle is the respective minimal ideal. In each case, it is clear that the socle is principal.

We further generalize Theorem~\ref{ring}.

\begin{corollary}
Let $R=GR(2^r,m)$ be a Galois ring with $r>1$. The two-sided principal ideal $(x)=\{0,x\}$, where
\begin{equation}
x=2^{r-1}(1+\omega+\dots+\omega^{m-1})(1+i+j+k)
\end{equation}
is the only minimal ideal of $\mathcal{H}(R)$.
\end{corollary}

\pf This follows from a tedious extension of the proof of Theorem~\ref{ring}.

We have a specific case which will be useful in a later example.

\begin{corollary}
The normalized homogeneous weight on $\mathcal{H}(\FF_2)$ is given as follows.
\[ w_{\textnormal{hom}}(x)=
\begin{cases}
0, & \text{if } x=0 \\
2, & \text{if } x=1+i+j+k\\
1, & \text{otherwise}
\end{cases}
\]
\end{corollary}

\subsection{Quaternion codes}\label{sect:results2}
Let $R=GR(p^r,m)$ be a Galois ring and let $\mathcal{H}$ be a quaternion ring over $R$. Let $B$ be a left $\mathcal{H}$-linear block code of length $n$. We consider the map $\tau:\mathcal{H}\rightarrow R^4$ defined by
\begin{equation}
\tau(x_0+x_1i+x_2j+x_3k)=(x_0,x_1,x_2,x_3)
\end{equation}
This map is a bijection and an additive homomorphism, and can be extended coordinate-wise to $\mathcal{H}^n$. It can be shown that $\tau(B)$ is a linear block code of length $4n$ over $R$. We equip $\tau(B)$ with a homogeneous distance metric with respect to the weight $w_{\textnormal{hom}}$ in (\ref{hom}) whose average value is given in (\ref{gamma}). Analogously, the right $\mathcal{H}$-linear block code $B'$ can be obtained with its corresponding image $\tau(B')$. However, even if the codes $B$ and $B'$ have the same generator matrix, $B$ and $B'$, and respectively $\tau(B)$ and $\tau(B')$, are not necessarily equal although they share the same distance properties.

The minimum Hamming weight of $B$ can be used to bound the minimum homogeneous distance of $\tau(B)$, as the next proposition shows. The proof below is analogous to that in \cite{sole}.

\begin{proposition}
Let $R=GR(p^r,m)$ be a Galois ring and let $\mathcal{H}$ be a quaternion ring over $R$. Let $B$ be a left $\mathcal{H}$-linear block code of length $n$ with minimum Hamming distance $d$, and $\tau(B)$ be the corresponding image code. Let $\delta$ denote the minimum homogeneous distance of $\tau(B)$ with respect to the homogeneous weight $w_{\textnormal{hom}}$. Then we have the bounds
\begin{equation}\label{rains}
\Gamma d \leq \delta \leq p^{m(r-1)}4d.
\end{equation}

\noindent Furthermore, let $\widetilde{\delta}$ be the minimum normalized homogeneous distance of $\tau(B)$ with respect to the normalized weight $\widetilde{w}_\textnormal{hom}$. Then we have the bounds

\begin{equation}
d \leq \widetilde{\delta} \leq \dfrac{p^m}{p^m-1}4d.
\end{equation}

\end{proposition}
\pf As $\Gamma$ is the minimum nonzero value of $w_{\textnormal{hom}}$, then $\Gamma d \leq \delta$. Furthermore, the minimum homogeneous distance $\delta$ is bounded above by $p^{m(r-1)}4n$. We have $\delta \leq w_{\textnormal{hom}}(\tau(x))$ for all $x \in B \setminus \{0\}$. Thus if $x$ is a codeword in $B$ of weight $d$, we then have $\delta \leq p^{m(r-1)} 4d$. The second bound follows immediately. \Q{\Box}

Let the minimum Hamming distance of $\tau(B)$ be denoted by $d_{\tau(B)}$. We say that $\tau(B)$ is Type $\alpha$ if
\begin{equation}\label{alpha}
\delta=p^{m(r-1)}d_{\tau(B)}.
\end{equation}
In particular, when $r=1$, the equality in (\ref{alpha}) reduces to $\delta=d_{\tau(B)}$, and when $m=1$, we have $\delta=p^{r-1}d_{\tau(B)}$ so that $\delta=2d_{\tau(B)}$ for codes over $\mathcal{H}(\ZZ_4)$.
\subsection{New examples}

A Fortran algorithm is written to verify the minimum distance of rate-$2/6$ codes. Note that this output is highly specific. However, the algorithm can be generalized to codes of any rate as desired and any number of variables in the generator matrix, at the cost of program runtime.

The algorithm first builds the quaternion ring, then iterates values for $x,y,z$ and builds a code for each iteration while computing for the weight of each codeword. It takes the minimum weight which corresponds to the minimum distance. Note that the codes are generally not two-sided.

\textbf{Example 1.} The left linear 3-quasi-cyclic rate-$2/6$ quaternion block code $B$ over $\mathcal{H}(\FF_3)$ is a free code with generator matrix

\begin{equation}
\begin{pmatrix}
1 & 1 & i & i & 1+i & 1+i\\
i & 1+i & 1+i & 1 & 1 & i
\end{pmatrix}.
\end{equation}

\noindent The code $B$ has 6561 codewords and minimum Hamming distance $d=5$. Its Hamming weight enumerator is given below.

\begin{equation}
W(x,y)=x^6 + 480xy^5 + 6080y^6
\end{equation}

\noindent The Hamming distance $d$ meets the Singleton upper bound as given in (\ref{singleton}) and therefore code $B$ is a maximum distance separable code. The code has minimum normalized homogeneous distance $75/16$. Its image $\tau(B)$ is a linear block code over $\FF_3$ of length $24$. Its minimum homogeneous distance $\delta$ is the same as its Hamming distance $d_{\tau(B)}$ which is $6$, and therefore by (\ref{alpha}), the code trivially is Type $\alpha$. Furthermore, the right linear 3-quasi-cyclic rate-$2/6$ quaternion block code with the above generator matrix has exactly the same distance properties. However, the two codes are not equal. For example, the codeword $\begin{pmatrix}
1+2k & 1+j+2k & i+j+2k & i+j & 1+i+j & 1+i+2k
\end{pmatrix}$ belongs to the left linear code but not to the right linear code. The image codes are also not equal.

\textbf{Example 2.} The 3-quasi-cyclic rate-$2/6$ quaternion block code $B$ over $\mathcal{H}(\FF_2)$ is a free code with generator matrix

\begin{equation}
\begin{pmatrix}
1 & 1 & i & i & 1+j & 1+j\\
i & 1+j & 1+j & 1 & 1 & i
\end{pmatrix}.
\end{equation}

\noindent The code $B$ has 256 codewords, minimum Hamming distance $d=4$ and minimum normalized homogeneous distance $4$. The code is nearly optimal with respect to (\ref{singleton}). Necessarily, the minimum homogeneous distance cannot be less than the minimum Hamming distance because they only differ at multiples of $1+i+j+k$. The code is two-sided since $\FF_2$ has characteristic $2$ and hence $\mathcal{H}(\FF_2)$ is commutative. Its image $\tau(B)$ is a linear block code over $\FF_2$ of length $24$. Its minimum homogeneous distance $\delta$ is the same as its Hamming distance $d_{\tau(B)}$ which is $8$, and therefore by (\ref{alpha}), the code trivially is Type $\alpha$.

\textbf{Example 3.} The left linear 3-quasi-cyclic rate-$2/6$ quaternion block code $B$ over $\mathcal{H}(\ZZ_4)$ is a free code with generator matrix

\begin{equation}
\begin{pmatrix}
1 & 1 & 2 & 2 & 1+4i & 1+4i\\
2 & 1+4i & 1+4i & 1 & 1 & 2
\end{pmatrix}.
\end{equation}

\noindent The code $B$ has 65536 codewords, minimum Hamming distance $d=4$ and minimum normalized homogeneous distance $4$. The code is nearly optimal with respect to (\ref{singleton}). Its image $\tau(B)$ is a linear block code over $\ZZ_4$ of length $24$. Its minimum homogeneous distance $\delta$ is the same as its Hamming distance $d_{\tau(B)}$ which is $8$. Furthermore, the right linear 3-quasi-cyclic rate-$2/6$ quaternion block code with the above generator matrix has exactly the same distance properties. However, the two codes are not equal. The image codes are also not equal.

\section{Conclusion and recommendations}
This paper dealt with one-sided linear block codes over non-commutative finite rings, in particular, the quaternion rings. By a suitable bijective map, these codes can be seen as linear block codes over finite commutative rings, specifically, the Galois rings. Furthermore, we have extended the notion of homogeneous weight to the quaternions and established certain bounds involving this weight, giving rise to good codes. It is also interesting to view a set of quaternions as a matrix code, and via lifting, to construct other quaternion codes that form a class similar to subspace codes in which an appropriate subspace metric can be applied. The authors recommend a more extensive review of the lattice structure of the quaternions over $GR(2^r,m)$ and the construction of other one-sided quaternion codes, in particular, over $GR(4,2)$. Finally, it is interesting to consider the equivalence between the left linear quaternion code and the right linear quaternion code and their corresponding image codes.

\section{Acknowledgement}
The authors gratefully acknowledge the members of the IMSP Coding Theory and Cryptography Research Cluster for helpful discussions.

\end{document}